\documentclass{IEEEcsmag}

\usepackage[colorlinks,urlcolor=blue,linkcolor=blue,citecolor=blue]{hyperref}

\usepackage{upmath}
\usepackage{multirow}

\jvol{XX}
\jnum{XX}
\paper{XX}
\jmonth{XX}
\jname{IEEE Software}
\pubyear{2020}

\begin{document}


\title{Combining Cognitive Styles Matters for Female Software Designers}

\author{C. Pretorius, M. Razavian, K. Eling, \& F. Langerak}
\affil{Eindhoven University of Technology, The Netherlands}

\markboth{}{Intuition and Rationality Both Matter for Software Design}

\begin{abstract}
Overcoming society’s complex problems requires novel solutions. Applying different cognitive styles can promote novelty when designing software aimed at these problems. Through an experiment with 80 software design practitioners, we found that female practitioners who had a preference for more than one cognitive style (intuition and rationality) produced the most novel software features of all participants.
\end{abstract}

\maketitle

\chapterinitial{There is consensus} in the software engineering community that practitioners sometimes rely on their intuition when designing software. Despite this, the emphasis in the software development process has generally been on promoting a rational cognitive style through rationalized processes, tools, and techniques. Meanwhile, the potential benefits of an intuitive style have been largely ignored \cite{Pretorius2018}. One such benefit of intuition is novelty \cite{Petervari2016}, which is crucial for tackling complex societal problems such as inequality, climate change, and health. 

To address this gap, we carried out an experiment in which software design practitioners with different cognitive styles designed software features for a mobile application to address a widespread health behavior problem. 

We found that female practitioners produced more novel software features than male practitioners, especially when they were both highly rational and highly intuitive.

Our study highlights the importance of considering and combining cognitive styles when designing new software features, but shows that female practitioners may uniquely benefit from combining intuitive and rational cognitive styles.

\section{WHY FEATURE NOVELTY?}

When designing software for a complex problem, software design practitioners, like product designers and requirements engineers, create new features to (partially) solve the problem. These practitioners tend to start off by sketching various ideas for a design on a whiteboard or piece of paper \cite{Mohanani2014}. They will then cycle back and forth between their understanding of the problem and their idea(s) for a potential feature, updating these concurrently as they go along. 

Nowadays, software solutions naturally lend themselves to addressing societal problems. However, the reality is that such problems demand substantial levels of novelty in software features \cite{Cropley2015}. 

\section{COGNITIVE STYLE AND GENDER MEET FEATURE NOVELTY}

Software design practitioners, like all people, have different cognitive styles. \textit{Cognitive style} describes differences in how people obtain, organize, and process information \cite{Aggarwal2019}. \textit{Intuition} and \textit{rationality} are two such cognitive styles. A practitioner designing a software feature through an intuitive style might do so quickly, and have a \textit{gut feeling} that their solution is the right one. Conversely, a practitioner using a rational style would arrive at a particular feature more slowly, justifying their solution in the context of available requirements. Both styles can be used by a practitioner at any time, in a particular order or even simultaneously. Still, all people tend to usually rely on one or both styles in a specific configuration \cite{Epstein1996}, known as their \textit{dispositional style}. 

Both intuition and rationality have been positively related to novelty. Intuition has been shown to result in more novel solutions through holistic information processing and promoting associative thinking \cite{Petervari2016}; the ``big picture". Rationality enables practitioners to assess \textit{details}, and to analytically compare potential solutions \cite{Epstein1996,Akinci2013}. Nevertheless, whether this is specifically true for software design practice remains to be seen.

Although dispositional style is not inherently gender-specific, it has been shown that the interaction between gender and job type can influence preference for intuition \cite{Akinci2013}. Given that female practitioners are often underrepresented in software engineering \cite{Vasilescu2012}, endure unique barriers to entering the field \cite{Wolff2020}, and are subject to a number of different biases \cite{Wang2019}, we were particularly curious about whether the novelty of software features designed by software design practitioners would vary based on their gender and dispositional style. When speaking of \textit{male} and \textit{female} in our study, we take gender to be a self-identification construct, which may or may not align with biology or presentation \cite{Burnett2016}.

Given these potential associations between cognitive style and feature novelty, and gender differences in style preference, our study investigated whether certain combination(s) of cognitive style and gender led to higher software feature novelty.

\section{STUDY DESIGN}

We conducted an experiment with practitioners to enable some control, while still maintaining real-world applicability. Practitioners, whose primary task involves high level design of features in any software engineering role, were recruited through the online platform \textit{Prolific}. Such participants are familiar with the complexity of the task, and comfortable with producing rough, wireframe-like sketches. First, participants took part in a feature design task. Afterwards, the same participants were randomly assigned to evaluate the novelty of ten features designed by others. 

We chose to focus on the health issue of obesity as our context, being a well-known issue that participants would at least be familiar with.

\subsection{Feature design task}

\begin{figure*}
\centerline{\includegraphics[width=32pc]{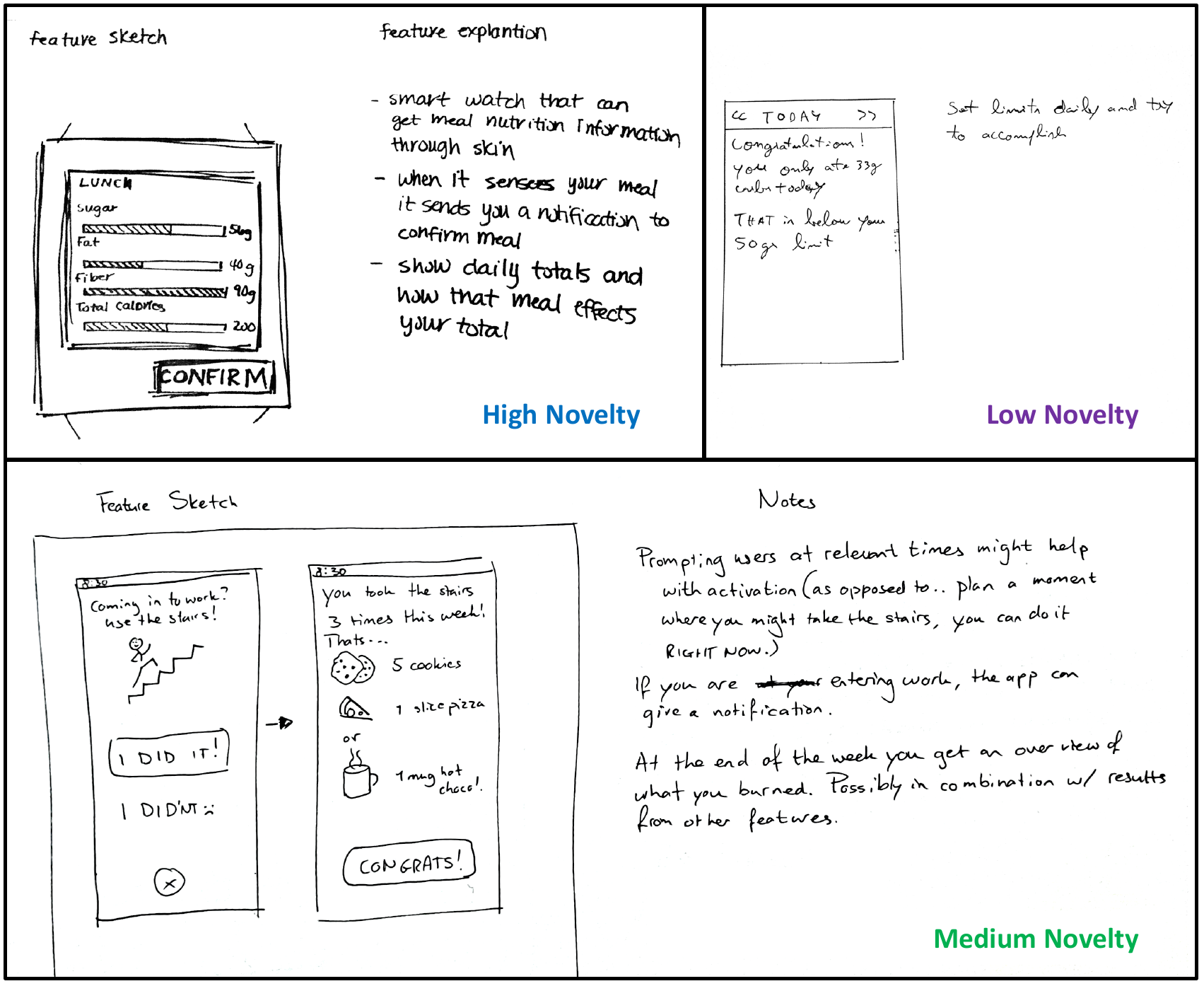}}
\caption{A selection of designed software features with high, low, and medium novelty scores respectively.}
\label{fig:features}
\end{figure*}

Participants were given an explanation of the problem, and instructed to design at least one feature for a mobile application. They were then given 15 minutes to sketch their software feature(s) on a piece of paper and provide suitable explanations, using a basic template as per \cite{Mohanani2014}. \textbf{Figure~\ref{fig:features}} presents a selection of the designed features.

Afterwards, we asked the participants to note which of their features, if they designed more than one, solved the problem best. The participants then photographed or scanned their features for upload. 

To measure participants’ dispositional cognitive style, we used the REI-10 (rational experiential inventory), which consists of five statements about participants’ use of intuition and five about their use of rationality \cite{Epstein1996}, measured on a 7-point scale from ``completely disagree" to ``completely agree." We dropped one item from the rationality scale that reduced the scale validity.

We collected participant's self-identified gender in the same section in which we asked control questions about work-relevant experience, industry role, age, and familiarity with the obesity problem. Participants were paid four English pounds for completing this part of the study.

\subsection{Feature evaluation task}

After completing the design task, participants were contacted again, and randomly formed into groups of five participants. Each group evaluated the same ten randomly selected features (always excluding their own). For each feature design sketch, participants were required to answer the question, ``How novel is this feature when compared with existing features from applications in the market?" Answers were recorded on a five-point scale ranging from ``not novel at all" to ``extremely novel". Participants were paid two English pounds for completing the evaluation of ten features.

To measure feature novelty, we calculated an average novelty score for each participant based on the five evaluations of their best (or only) feature.

\subsection{The sample}

110 practitioners had their top-rated feature evaluated. This was reduced to 80 following data cleaning. 26.25\% of the participants were female, and
73.75\% were male. 23.8\% of the participants had a high preference for intuition, but low for rationality; 22.5\% had a high preference for rationality, but low for intuition; 22.5\% high preference for both; and 31.2\% for neither cognitive style. Participants' professional design experience varied from less than one year to more than 20 years, with the average being 5.44 years.

\subsection{Data analysis}

We used hierarchical moderated regression analysis to determine whether gender, intuitive style or rational style could significantly account for differences in feature novelty in isolation and when taken together. For this purpose, the three variables were included individually and in all possible two-way and three-way combinations in our model. To account for other influences, we initially included the number of features designed, experience, and age of participants in the model, but these did not correlate with feature novelty. With an R-squared value of .196 and a coefficient F-value of 2.509, the variables included in the final model accounted for 19.6\% of the variance in feature novelty among participants (constant 2.510). 

\section{HOW COGNITIVE STYLE AND GENDER EXPLAIN FEATURE NOVELTY}

\subsection{Cognitive style alone does not matter}

We found that cognitive style was unrelated to feature novelty on its own. Neither a more intuitive nor a more rational dispositional style per se led participants to design a software feature of higher novelty.

\subsection{Female practitioners create more novel features} 

Gender, in contrast, was positively associated with feature novelty. We found that the female practitioners in our experiment produced more novel software features than the male practitioners did. 

\subsection{Cognitive style matters for female practitioners} 

Cognitive style and gender taken together are also positively related to feature novelty. Female practitioners with a higher intuitive preference designed significantly more novel software features. Additionally, we found that female practitioners produced the most novel features when they had a preference for both intuition and rationality. 

\begin{figure*}
\label{process}
\centerline{\includegraphics[width=26pc]{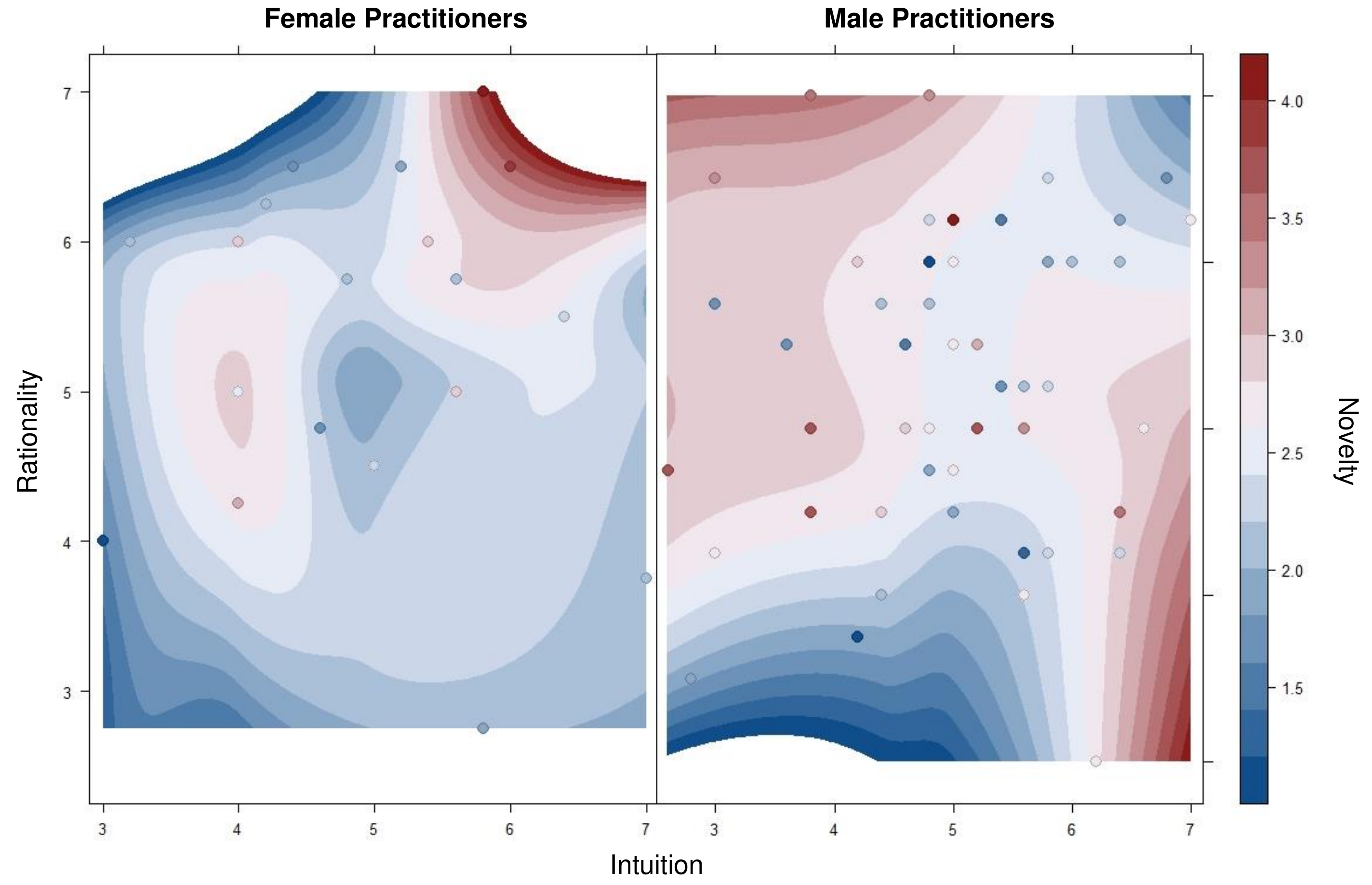}}
\caption{Graphical representation of the relationship between cognitive style and software feature novelty, separated by gender. Red regions show higher novelty, whereas blue regions show lower novelty.}
\label{fig:heatmap}
\end{figure*}

The two heatmaps in \textbf{Figure~\ref{fig:heatmap}} illustrate the relationships between intuition, rationality, and feature novelty for male and for female practitioners. It is important to keep in mind that only the high intuition, and high intuition with high rationality portions in the female practitioner of the regression model, are statistically significant.

\section{DISCUSSION AND KEY TAKEAWAYS}

Our study shows that both cognitive styles (intuition and rationality) as well as gender matter for software feature novelty. The way that they matter leads to several important takeaways from our study.

First, since gender was positively associated with software feature novelty in our study, it is crucial to further investigate the role of female practitioners in software design activities. Perhaps software teams could benefit from involving women specifically, but this needs to be empirically established.

Second, since neither cognitive style was positively related to novelty on its own, it does not make sense to lean on a single cognitive style, independent of other factors, to design novel software features. Previous research and practice in software engineering has generally prescribed the use of rationality either explicitly (e.g. through design reasoning techniques \cite{Tang2018}) or implicitly (e.g. by imposing structured development methods and lifecycle models \cite{Ralph2018}). Focusing entirely on rationality or intuition is not supported by our study. Instead, other factors need to be considered, particularly the gender of the practitioner.

Indeed, we found that gender has implications for choosing a cognitive style in software feature design. Female practitioners should not be discouraged from making use of intuition exclusively, or combining rationality with intuition, when designing software features. In practice, intuition can even be promoted through behaviors such as brainstorming and sketching to intuitively come up with potential solutions, as well as giving female practitioners \textit{incubation} time (i.e., distraction from consciously considering the problem) after being exposed to a problem situation \cite{Gilhooly2012}. 

Currently, we cannot draw certain conclusions from the male portion of our regression model. However, it is possible that male practitioners with a dispositional preference for either intuition or rationality design more novel software features than their male peers in contexts other than our study. This also raises the question of whether forcing a rational style among intuitively strong male practitioners is actually beneficial for these practitioners.

\section{LIMITATIONS AND FUTURE WORK}

To our knowledge, this is the first study that investigates the relationship between the combination of cognitive style and gender, and software-related performance outcomes. We hope that it will encourage further work on this important subject. However, in our study design, we have solely considered a \textit{black-box} approach to the problem, which does not consider \textit{how} male and female practitioners might design software features differently. Such \textit{white-box} studies, particularly qualitative design studies, are imperative for understanding these differences in practice. Perhaps the differences we found can be explained by the pressure sometimes experienced by female practitioners to prove themselves \cite{Blincoe2019}, as an example.

The black-box nature of our study leads to two further potential limitations. First, although we controlled for many extraneous variables, there are likely other variables, such as self-confidence \cite{Vasilescu2012}, that we did not control for. Second, although our sample consists of software design practitioners from many different geographic locations and roles in industry, it is possible that our sample is not perfectly representative. 

Finally, our study focused on the individual level. Although some aspects of studying individual practitioners can be applied to the team level, researchers should also investigate the novelty of software designed by teams. Teams can differ in terms of cognitive style and gender representation. Interactions between individual practitioners based on these differences could have unique consequences for software novelty, perhaps through issues like groupthink and power dynamics invading or supporting the group context.

\bibliographystyle{IEEEtran}
\bibliography{bib}

\begin{thebibliography}{10}
\providecommand{\url}[1]{#1}
\csname url@samestyle\endcsname
\providecommand{\newblock}{\relax}
\providecommand{\bibinfo}[2]{#2}
\providecommand{\BIBentrySTDinterwordspacing}{\spaceskip=0pt\relax}
\providecommand{\BIBentryALTinterwordstretchfactor}{4}
\providecommand{\BIBentryALTinterwordspacing}{\spaceskip=\fontdimen2\font plus
\BIBentryALTinterwordstretchfactor\fontdimen3\font minus
  \fontdimen4\font\relax}
\providecommand{\BIBforeignlanguage}[2]{{%
\expandafter\ifx\csname l@#1\endcsname\relax
\typeout{** WARNING: IEEEtran.bst: No hyphenation pattern has been}%
\typeout{** loaded for the language `#1'. Using the pattern for}%
\typeout{** the default language instead.}%
\else
\language=\csname l@#1\endcsname
\fi
#2}}
\providecommand{\BIBdecl}{\relax}
\BIBdecl

\bibitem{Pretorius2018}
C.~Pretorius, M.~Razavian, K.~Eling, and F.~Langerak, ``Towards a dual
  processing perspective of software architecture decision making,'' in
  \emph{IEEE 15th International Conference on Software Architecture Companion
  (ICSA-C 2018)}.\hskip 1em plus 0.5em minus 0.4em\relax Piscataway: IEEE,
  2018, pp. 48--51.

\bibitem{Petervari2016}
J.~P{\'{e}}terv{\'{a}}ri, M.~Osman, and J.~Bhattacharya, ``The role of
  intuition in the generation and evaluation stages of creativity,''
  \emph{Frontiers in Psychology}, vol.~7, no. 1420, pp. 1--12, 2016.

\bibitem{Mohanani2014}
R.~Mohanani, P.~Ralph, and B.~Shreeve, ``Requirements fixation,'' in
  \emph{Proceedings of the 36th International Conference on Software
  Engineering (ICSE '14)}.\hskip 1em plus 0.5em minus 0.4em\relax Hyderabad,
  India: ACM, 2014, pp. 895--906.

\bibitem{Cropley2015}
D.~H. Cropley, \emph{Creativity in engineering: Novel solutions to complex
  problems}.\hskip 1em plus 0.5em minus 0.4em\relax London, UK: Academic Press,
  2015.

\bibitem{Aggarwal2019}
I.~Aggarwal and A.~W. Woolley, ``Team creativity, cognition, and cognitive
  style diversity,'' \emph{Management Science}, vol.~65, no.~4, pp. 1586--1599,
  2019.

\bibitem{Epstein1996}
S.~Epstein, R.~Pacini, V.~Denes-Raj, and H.~Heier, ``Individual differences in
  intuitive–experiential and analytical–rational thinking styles.''
  \emph{Journal of Personality and Social Psychology}, vol.~71, no.~2, pp.
  390--405, 1996.

\bibitem{Akinci2013}
C.~Akinci and E.~Sadler-Smith, ``Assessing individual differences in
  experiential (intuitive) and rational (analytical) cognitive styles,''
  \emph{International Journal of Selection and Assessment}, vol.~21, no.~2, pp.
  211--221, 2013.

\bibitem{Vasilescu2012}
B.~Vasilescu, A.~Capiluppi, and A.~Serebrenik, ``Gender, representation and
  online participation: A quantitative study of {{StackOverflow}},''
  \emph{Proceedings of the 2012 ASE International Conference on Social
  Informatics, SocialInformatics 2012}, pp. 332--338, 2012.

\bibitem{Wolff2020}
A.~Wolff, A.~Knutas, and P.~Savolainen, ``What prevents {{F}}innish women from
  applying to software engineering roles? {{A}} preliminary analysis of survey
  data,'' in \emph{Software Engineering Education and Training (ICSE-SEET
  '20)}.\hskip 1em plus 0.5em minus 0.4em\relax New York, NY, USA: ACM, 2020.

\bibitem{Wang2019}
Y.~Wang and D.~Redmiles, ``Implicit gender biases in professional software
  development: An empirical study,'' \emph{Proceedings of the 41st
  International Conference on Software Engineering: Software Engineering in
  Society (ICSE-SEIS '19)}, pp. 1--10, 2019.

\bibitem{Burnett2016}
M.~Burnett, S.~Stumpf, J.~Macbeth, S.~Makri, L.~Beckwith, I.~Kwan, A.~Peters,
  and W.~Jernigan, ``Gendermag: A method for evaluating software's gender
  inclusiveness,'' \emph{Interacting with Computers}, vol.~28, no.~6, pp.
  760--787, 2016.

\bibitem{Tang2018}
A.~Tang, F.~Bex, C.~Schriek, and J.~M.~E. van~der Werf, ``Improving software
  design reasoning – a reminder card approach,'' \emph{Journal of Systems and
  Software}, vol. 144, no. February, pp. 22--40, 2018.

\bibitem{Ralph2018}
P.~Ralph, ``The two paradigms of software development research,'' \emph{Science
  of Computer Programming}, vol. 156, no. May, pp. 68--89, 2018.

\bibitem{Gilhooly2012}
K.~J. Gilhooly, G.~J. Georgiou, J.~Garrison, J.~D. Reston, and M.~Sirota,
  ``Don't wait to incubate: Immediate versus delayed incubation in divergent
  thinking,'' \emph{Memory and Cognition}, vol.~40, no.~6, pp. 966--975, 2012.

\bibitem{Blincoe2019}
K.~Blincoe, O.~Springer, and M.~R. Wr{\'{o}}bel, ``Perceptions of gender
  diversity's impact on mood in software development teams,'' \emph{IEEE
  Software}, vol.~36, no.~5, pp. 51--56, 2019.

\end{thebibliography}

\begin{IEEEbiography}{Carianne Pretorius}{\,}is a doctoral candidate in the School of Industrial Engineering at Eindhoven University of Technology in the Netherlands. Her research interests include software design decision making, cognition, and human aspects of software engineering. Pretorius received her M.A. in Socio-Informatics \textit{cum laude} from Stellenboch University in South Africa. Contact her at c.pretorius@tue.nl.
\end{IEEEbiography}

\begin{IEEEbiography}{Maryam Razavian}{\,}is an assistant professor in industrial engineering at Technical University of Eindhoven. Her research interests include software design reasoning, human aspects of software design, software architecture, and service orientation. Razavian received her Ph.D. in Computer Science from VU University Amsterdam. She's a member of ACM and IEEE. Contact her at m.razavian@tue.nl.
\end{IEEEbiography}

\begin{IEEEbiography}{Katrin Eling}{\,}is an assistant professor of new product development in the School of Industrial Engineering at Eindhoven University of Technology in the Netherlands. She received her Ph.D. from the same school, has an M.Sc. in strategic product design from Delft University of Technology, and a diploma in industrial design from the University of Wuppertal, Germany. Dr. Eling's research focuses on the successful management of the front end of innovation. Contact her at k.eling@tue.nl.
\end{IEEEbiography}

\begin{IEEEbiography}{Fred Langerak}{\,}is professor of product development and management in the Innovation, Technology Entrepreneurship \& Marketing Group of the School of Industrial Engineering at Eindhoven University of Technology in the Netherlands. He has a M.Sc. and Ph.D. from the Erasmus School of Economics. His research focuses on managerial interventions to improve the process of conceiving, designing, developing, and bringing new products to market, and managing these products post‐launch. Contact him at f.langerak@tue.nl.
\end{IEEEbiography}

\end{document}